\documentclass[runningheads]{llncs}
\usepackage{graphicx}
\usepackage{subfigure}
\usepackage{color}
\usepackage{multirow}
\usepackage{cite}
\newtheorem{myDef}{Definition}
\begin{document}
	\title{Exploring EOSIO via Graph Characterization}
	%
	%
	\author{Yijing Zhao\inst{1,2,3}\and
	Jieli Liu\inst{1,2}\and
	Qing Han\inst{1,2}\and
	Weilin Zheng\inst{1,2}\and
	Jiajing Wu\inst{1,2}}

	\authorrunning{Y. Zhao et al.}
	%
    \institute{
    School of Data and Computer Science, Sun Yat-sen University, Guangzhou 510006, China\and
    National Engineering Research Center of Digital Life, Sun Yat-sen University, Guangzhou, China\and
	School of Electronics and Communication Engineering, Sun Yat-sen University, Guangzhou 510006, China\\
	\email{\{zhaoyj53, liujli7, hanq25, zhengwlin\}@mail2.sysu.edu.cn}\\ \email{\{wujiajing\}@mail.sysu.edu.cn}}
	\maketitle              
	\begin{abstract}
	Designed for commercial decentralized applications (DApps), EOSIO is a Delegated Proof-of-Stake (DPoS) based blockchain system. It has overcome some shortages of the traditional blockchain systems like Bitcoin and Ethereum with its outstanding features (e.g., free for usage, high throughput and eco-friendly), and thus becomes one of the mainstream blockchain systems. Though there exist billions of transactions in EOSIO, the ecosystem of EOSIO is still relatively unexplored.
	To fill this gap, we conduct a systematic graph analysis on the early EOSIO by investigating its four major activities, namely account creation, account vote, money transfer and contract authorization.
	We obtain some novel observations via graph metric analysis, and our results reveal some abnormal phenomenons like voting gangs and sham transactions.
	
	\end{abstract}
	
	%
	
	
	%
	%
	
	\section{Introduction}
	Recent years, blockchain technology has become a buzzword and aroused a great deal of interests among researchers, developers and investors. Among the blockchain systems, Ethereum is the largest one that supports smart contracts. However, it suffers from high transaction-confirmation latency and low throughput problems since the employ of Proof-of-Work (PoW) consensus protocol.
	And it gradually becomes unable to meet the demand of the rapid development of decentralized applications (DApps), which need higher scalability and quicker response for transaction confirmations.
	Based on the Delegated Proof-of-Stake (DPoS) consensus, a new platform EOSIO provides a solution for these problems. Built for commercial DApps, EOSIO has some outstanding features like free for usage, high throughput and eco-friendly, having attracted much attention. Especially, the number of transactions in EOSIO has reached more than four billion within two years, which witnesses the prosperity of EOSIO.
	
	There are some studies about the performance \cite{xu2018eos}, security \cite{238604} \cite{he2020security}, transaction data analysis \cite{huang2020characterizing} \cite{zheng2020xblock} of EOSIO. For example, Xu et al. \cite{xu2018eos} presented a thorough analysis on EOSIO from the perspectives of architecture, performance, and economics. Lee et al. \cite{238604} conducted the first study to analyze the security and possible attacks of EOSIO. Huang et al. \cite{huang2020characterizing} characterized the activities in EOSIO and developed techniques for detecting bots and fraudulent activities based on their insights. Zheng et al. \cite{zheng2020xblock} provided an overview of up-to-date on-chain data of EOSIO. However, existing studies investigating the EOSIO ecosystem from a graph analysis perspective are limited. And more in-depth analyses are needed to discover the user behaviors and understand EOSIO.
	
	
	In this paper, we utilize graph analysis to explore the characteristics of the early EOSIO by investigating four kinds of user activities, namely account creation, account vote, money transfer and contract authorization. Firstly, according to these four kinds of activities in the first 15 million blocks, we construct the account creation graph (ACG), money transfer graph (MTG), account vote graph (AVG) and contract authorization graph (CAG) as weighted directed graphs. Secondly, we conduct an analysis on these graphs by measuring some graph metrics such as degree distribution, clustering coefficient, connected component, etc. Finally based on the investigation results, we discover some interesting insights about the EOSIO ecosystem, which would help people understand the user activities in the early EOSIO.
	
	\subsection{Related Work}
	
	Graph analysis assists people to understand the relationship between objects in complex systems. In 2012, Reid and Harrigan \cite{reid2013analysis} first modeled the Bitcoin transaction data with graph representations. By combining some external information, they investigated a theft case of Bitcoin with flow analysis.
	Up to now, many researchers have conducted graph analysis on blockchain transaction data. Existing work on blockchain transaction graph analysis can be divided into describing the graph properties via some metrics, and conducting data mining tasks on graph-structure data. The former can give us insights into the blockchain systems and how their transaction graphs form and develop, while the latter mainly investigates some data mining tasks such as de-anonymizing the accounts \cite{meiklejohn2013fistful}, detecting illicit activities \cite{chen2019exploiting, chen2019market}, etc. And our work focuses on the former one.
	
	There are many studies on investigating the blockchain transaction graph with graph metrics. For example, Lischke and Fabian \cite{lischke2016analyzing} examined the Bitcoin transaction graph and economy during the first four years, and this analysis revealed the business distribution as well as transaction distribution across countries, and investigated the small world phenomenon in some subgraphs.
	Chen et al. \cite{chen2018understanding} analyzed three major activities (money transfer, account creation and contract invocation) in Ethereum via graph analysis, and they discovered some new observations which help people have a full understanding of Ethereum.
	Motamed and Bahrak \cite{pasha2019quantitative} investigated the graph properties of five kinds of cryptocurrencies and compared the evolution of these properties between different cryptocurrencies.
	Since EOSIO is a newly emerging blockchain system, studies on graph properties analysis related to EOSIO are few. Huang et al. \cite{huang2020characterizing} analyzed the money transfer, account creation and contract invocation activities of EOSIO, and further developed techniques to detect bots and fraudulent activities. However, our study are focused on characterizing four main activities (namely, account creation, account vote, money transfer and contract authorization) in the early EOSIO with graph property analysis, and we provide a deeper insight into the graph properties.
	
	\subsection{Contribution}
	In summary, we investigate the four major behaviors in EOSIO by conducting a graph analysis. Our major contributions are listed as follows:
	\begin{enumerate}
			\item[(1)] To the best of our knowledge, our work is the first systematic and comprehensive research on the early EOSIO via analyzing four major activities, namely account creation, account vote, money transfer and contract authorization.
			\item[(2)] We construct four graphs for the four major activities in EOSIO, by measuring some graph metrics such as clustering coefficient, assortativity and so on, we obtain some interesting insights.
			\item[(3)] We observe some abnormal phenomenons like voting gangs and sham transactions in EOSIO during our analysis, which helps the supervision enhancement of EOSIO.
	\end{enumerate}
	
	The remaining sections are organized as follows. We introduce some background knowledge of EOSIO in Section \ref{Background}. Then, We detail the procedure of data collection in Section \ref{DataCollection}. Next, we conduct graph construction and graph analysis in Section \ref{GraphAnalysis}, where we list some analysis results. And finally, we conclude this paper in Section \ref{Conclusion}.
	
	\section{Background}\label{Background}
	This section introduces some background knowledge of EOSIO related to the following research. More details about the operation of  EOSIO can be found in its white paper \cite{TechnicalWhitePaper}.
	
	\textbf{What's EOSIO?} Released in June, 2018, EOSIO is a DPoS-based blockchain system designed for building commercial DApps. In DPoS, only 21 block producers are in charge of transaction verification and block production. These block producers are chosen by the vote from the token holders in EOSIO, which can guarantee the fairness and choose the most trusted 21 block producers. Compared with traditional blockchain systems, the DPoS-based EOSIO has much higher throughput per second (tps) that can generate a block with an average of 0.5 seconds. Like bitcoin in the Bitcoin system and Ether in Ethereum, the most common currency token in EOSIO is named EOS. Besides, EOSIO can also support Turing-complete smart contracts, and it provides a more complete smart contract ecosystem. The complied bytecode of each contract is executed in EOSIO's WebAssembly-based virtual machine (EOSVM). Unlike many blockchain systems that use gas mechanism to solve halting problem \cite{turing1938computable}, EOSIO is resource constrained (i.e., limiting the RAM, CPU and bandwidth). The resources can be obtained by token mortgage in EOSIO, and it is almost free for users because of the resource supplied from many DApps.
	
	\textbf{Transactions and actions.} In EOSIO, a block can contain multiple transactions and each transaction is made up of one or more actions. An action is an invocation of a contract that represents an operation in the system. There are mainly three types of actions, namely calling action, inline action and deferred action. A calling action represents a contract invocation from a user while an inline action represents an invocation triggered by a contract. And a deferred action is an action being scheduled to execute in a future transaction. Once a calling action or an inline action fails, the transaction would be rolled back, while the failure of a deferred action only affects the scheduled transaction. It's worth mentioned that the detailed information of inline actions does not be packaged into transactions, which brings challenges in transaction data acquisition.
	
	\textbf{Accounts and permissions.} Different from Ethereum, the identity of an account is a unique string containing up to 12 characters, but not the public key. An account can only be created by an existing account in EOSIO, except the initial account in EOSIO named \textit{eosio}. Each account can deploy only one contract on itself through the \textit{setcode} interface of \textit{eosio}, and can delete this contract by setting the contract code empty with the same interface. The basic actions are completed by interfaces provided from system accounts. For example, new accounts can be created by the \textit{newaccount} interface of \textit{eosio}, and the EOS transfer operation can be executed by the \textit{transfer} interface of \textit{eosio.token}. When invoking a contract, the related accounts should delegate appropriate permissions for the execution, namely assign specific public/private keys and grant privileges to this action. The permissions are generally divided into owner permission and active permission, where the owner permission is the highest level of permission and it is designed for cold storage, and the active permission can perform all operations except changing the owner.
	\section{Data Collection} \label{DataCollection}
	We collect all transaction data of the first 15 million blocks, which includes about 3 months transaction data from the launch of EOSIO on June 6, 2018.
	
	Since the large volume of transaction data, it is infeasible to obtain all transaction data we need by directly crawling them from blockchain explorer. We first utilize Nodeos, an EOSIO client provided by the official EOSIO development team, to synchronize the on-chain data. To speed up this process, we download blocks from some EOSIO backup service provider firstly and then start Nodeos from a certain specified block. Then the transaction information can be obtained through the RPC interface of Nodeos.
	
	However, the details of inline actions do not be recorded in on-chain data. To address this issue, we replay all transactions and utilize the action trace data, which is generated in EOSVM and records the detailed run-time information of actions. By calling the \textit{history\_file\_plugin} interface provided by the EOSIO development team, we obtain the action trace data. Then, we extract the actions of activities including account creation, account vote, money transfer and contract authorization from the raw data. Noting that the contract authorization activity refers to the permission delegating behavior for contracts belonging to common users rather than system users like \textit{eosio.token}. After the representation simplification of the raw data, we obtain the data that can be directly applied to graph analysis.
	
	The statistics of actions for the four activities is shown in Table \ref{Table:Proportion}. As we can see, most of these activities are accomplished with calling actions. The proportion of calling actions in the actions of account creation and account vote are not surprising, since EOSIO provides interfaces for these activities from system accounts. Especially, we can conclude that the voting behaviors are relatively transparent in the early EOSIO because most of them are accomplished with calling actions whose records are public accessible in blockchain. Though money transfer activity can be conducted by directly calling the interface provided by \textit{eosio.token} through calling actions, the proportion of inline actions in money transfer activity is relatively high, which means smart contracts are widely used in setting specific transaction rules by users. For contract authorization, it also has a small proportion of inline actions, and this phenomenon is caused by the self\-invocation between contracts.
	
	\begin{table}
		\renewcommand{\arraystretch}{1.2}
		\caption{Statistics of actions.}
		\centering
		\setlength{\tabcolsep}{1.2mm}{
			\begin{tabular}{|c|c|c|}
				\hline
				\centering
				Activity & Calling action (proportion) & Inline action (proportion) \\
				\hline
				Account creation & 299,178 (99.053\%) & 2,860 (0.947\%) \\
				\hline
				Account vote & 129,800 (99.997\%) & 4 (0.003\%)\\
				\hline
				Money transfer & 4,557,498 (52.857\%) & 4,064,790 (47.143\%)\\
				\hline
				Contract authorization & 329,012,816 (99.241\%) & 2,515,729 (0.759\%) \\
				\hline
		\end{tabular}}
		\label{Table:Proportion}
	\end{table}
	
	\section{Graph Analysis} \label{GraphAnalysis}
	
	In this section, we explore the account creation, account vote, money transfer and contract authorization activity in EOSIO through graph analysis. By investigating several graph metrics, We obtain some interesting insights as follows:\\
	\textbf{Insight 1:} In the early EOSIO, some accounts participate in transactions for testing or experiencing the platform.\\
	\textbf{Insight 2:} Some risks exist in EOSIO, like the voting gangs in which the members vote for each other and the observing abnormal account for pressure testing.\\
	\textbf{Insight 3:} EOSIO may exist spam transactions in its billions of transactions.
	
	\subsection{Account Creation}	
	A new account is created by an existing account in EOSIO. To model the account creation activity, we define the account creation graph (ACG) as follow:

	\begin{myDef}[ACG] An account creation graph (ACG) is a directed graph $G=(V,E)$, where $V$ is the set of nodes representing accounts in EOSIO and $E$ is the set of edges, in which each edge $(v_i,v_j)$, $v_i,v_j\in V$ represents the account creation relationship that $v_i$ create $v_j$.
	\end{myDef}

	\begin{figure*}[htbp]
	\centerline{
		\includegraphics[scale=0.10]{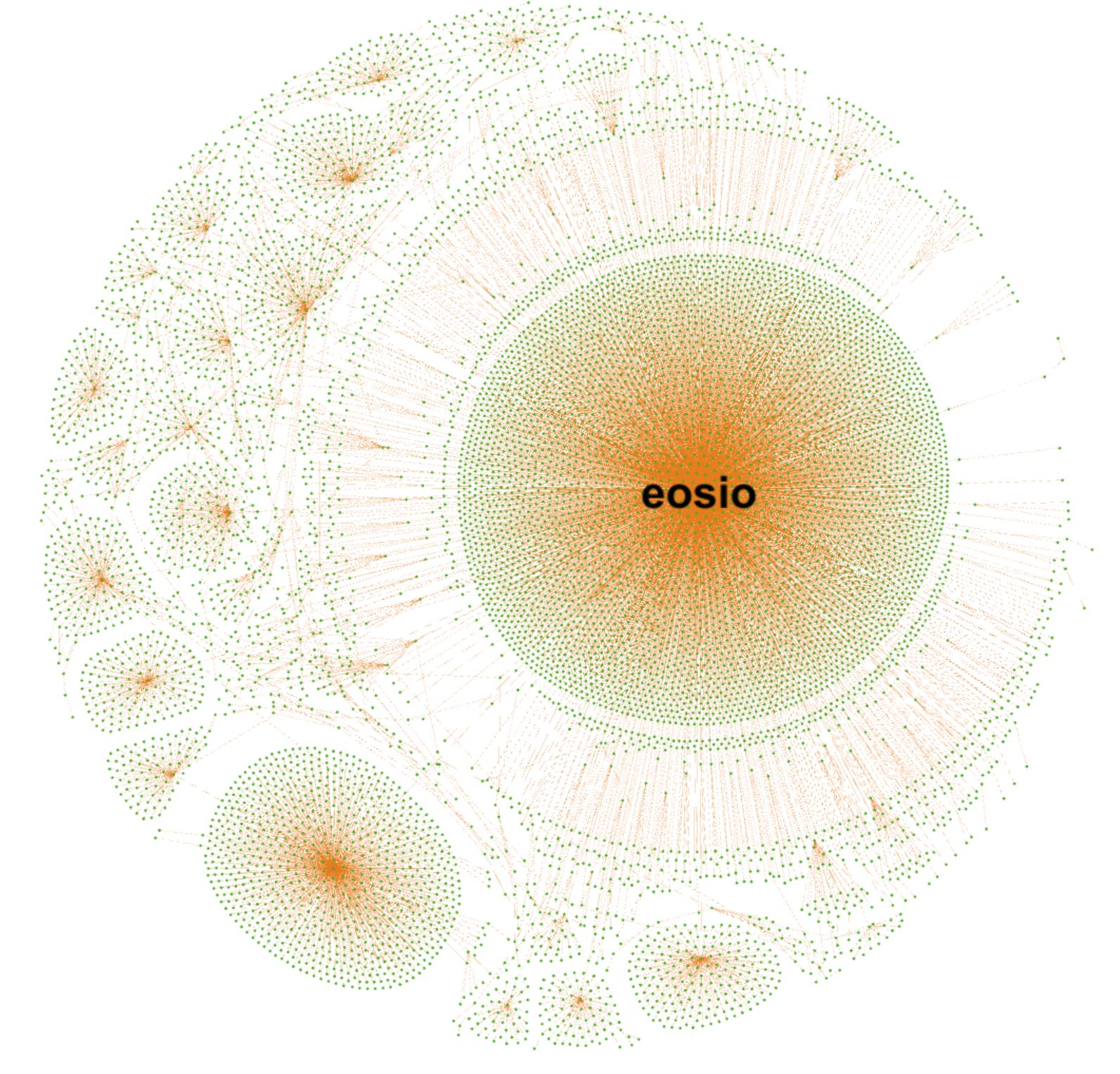}
	}
	\caption{The visualization of ACG.}
	\label{fig:ACGVisualization}
	\end{figure*}
	
	There are total 302,039 nodes and 302,038 edges in the constructed ACG. We can see that the edge number is the same as the action number of account creation in Table \ref{Table:Proportion}, and the node number is one more than the edge number, which implies that each node except the initial system account \textit{eosio} can be created once by its father account. Since a father account must be an existing account in EOSIO, the ACG is a tree-like graph with no circle.
	
	We then visualize ACG by randomly selecting 8,000 nodes and applying union-find algorithm \cite{Kozen1992} to find out all connecting paths to the ancestor for each node. The visualization result of the graph including all selected edges via union-find algorithm is shown in Fig. \ref{fig:ACGVisualization}. It is obviously that the graph is in a tree-like structure where most of the nodes have no outdegree, and every node has a connecting path to \textit{eosio}, the ancestor of other nodes in the graph. In addition, there are several clusters in the graph and the cluster which includes \textit{eosio} is the biggest one. 

	\begin{figure}[htbp]
		\centering
		\subfigure[Degree distribution]{%
			\includegraphics[scale=0.275]{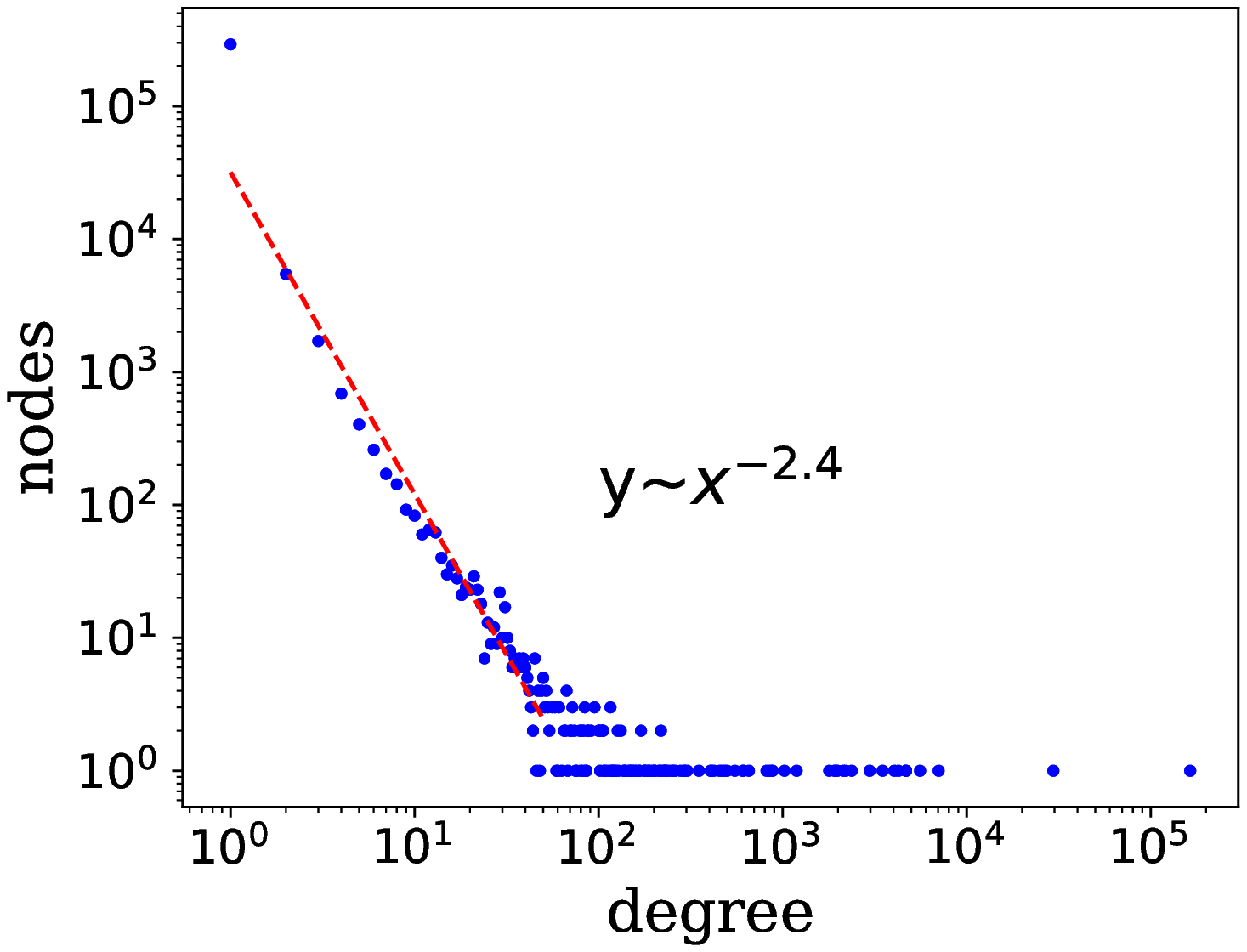}
		}
		\subfigure[Outdegree distribution]{%
			\includegraphics[scale=0.28]{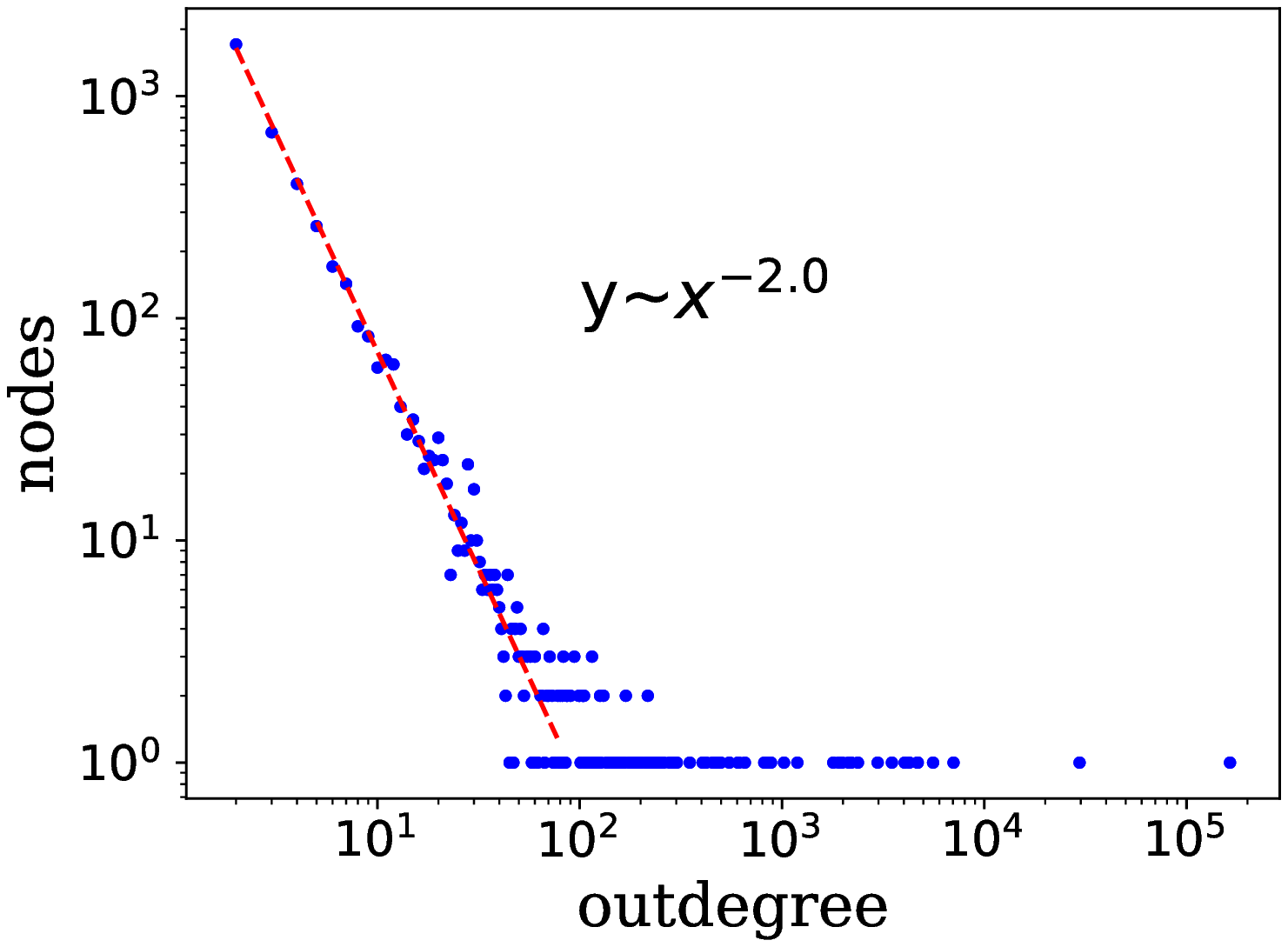}
		}
		\centering
		\caption{Degree/Outdegree distributions of ACG}
		\label{fig:ACGdegree}
	\end{figure}
	
	Fig. \ref{fig:ACGdegree} displays the degree distribution and outdegree distribution of ACG. Both of them satisfy the power law distribution and have a long tail.
	These distributions indicate that there are a few accounts creating many accounts with large degree.
	Besides, since the initial system account \textit{eosio} have no indegree, and the indegree of other accounts is 1 because they can only be created once, we do not present the indegree distribution of ACG.

	\begin{table}
		\caption{Metrics of Graphs}
		\begin{tabular}{|c|c|c|c|c|c|c|c|c|}
			\hline
			\centering
			\multirow{2}*{Graph} & \multirow{2}*{Cluster} & \multirow{2}*{Assortativity} & Number of & Largest & Number of & Largest & Largest & Smallest\\
			&&& SCC & SCC & WCC & WCC & diameter & diameter\\
			\hline
			ACG & 0 & / & 302,039 & 1 & 1 & 302,039 & 20 & 20 \\
			\hline
			AVG & 0.066 & -0.221 & 28,750 & 18 & 3 & 28,766 & 6 & 1 \\
			\hline
			MTG & 0.259 & -0.338 & 149,536 & 55,238 & 1 & 204,841 & 6 &  6 \\
			\hline
			CAG & 0.086 & -0.160 & 35,462 & 3 & 223 & 35,086 & 10 & 0\\
			\hline
		\end{tabular}
		\label{table:Metrics of Graphs}
	\end{table}

	Table \ref{table:Metrics of Graphs} shows some graph metrics of ACG, including the values of the clustering coefficient, assortativity coefficient, number of SCC/WCC, number of nodes in the largest SCC/WCC, and the largest/smallest diameter of WCC.
	We can see that the clustering coefficient is 0, because there is no account creation relationship between two accounts created by the third node. We have mentioned the fact that ACG is in a directed tree-like structure where each account can only be created by an existing account. Hence the number of nodes in the largest SCC is 1 and all nodes are in the largest WCC, which match our expectations. The largest diameter of WCC is equal to the smallest one since ACG has one WCC, and the value of diameter indicates that the height of the tree is 20.

	\subsection{Account Vote}
	In EOSIO, each account can vote for others to choose the 21 block producers. To investigate the relationship between voters and candidates, we define the account vote graph (AVG) as follow:
%
%

	\begin{myDef}[AVG] An account vote graph (AVG) is a directed weighted graph $G=(V,E,W)$, where $V$ denotes the set of nodes representing accounts enrolling in EOSIO's voting activity, $E$ denotes the set of edges, in which each edge $(v_i,v_j)$, $v_i,v_j\in V$ represents that $v_i$ votes for $v_j$, and there is a mapping function ${\varphi}: E\rightarrow W$ that maps a weight from the edge attribute set $W$ for each edge, which represents the corresponding voting times.
	\end{myDef}

	\begin{figure*}[htbp]
	\centerline{
		\includegraphics[scale=0.12,trim=20 20 20 10]{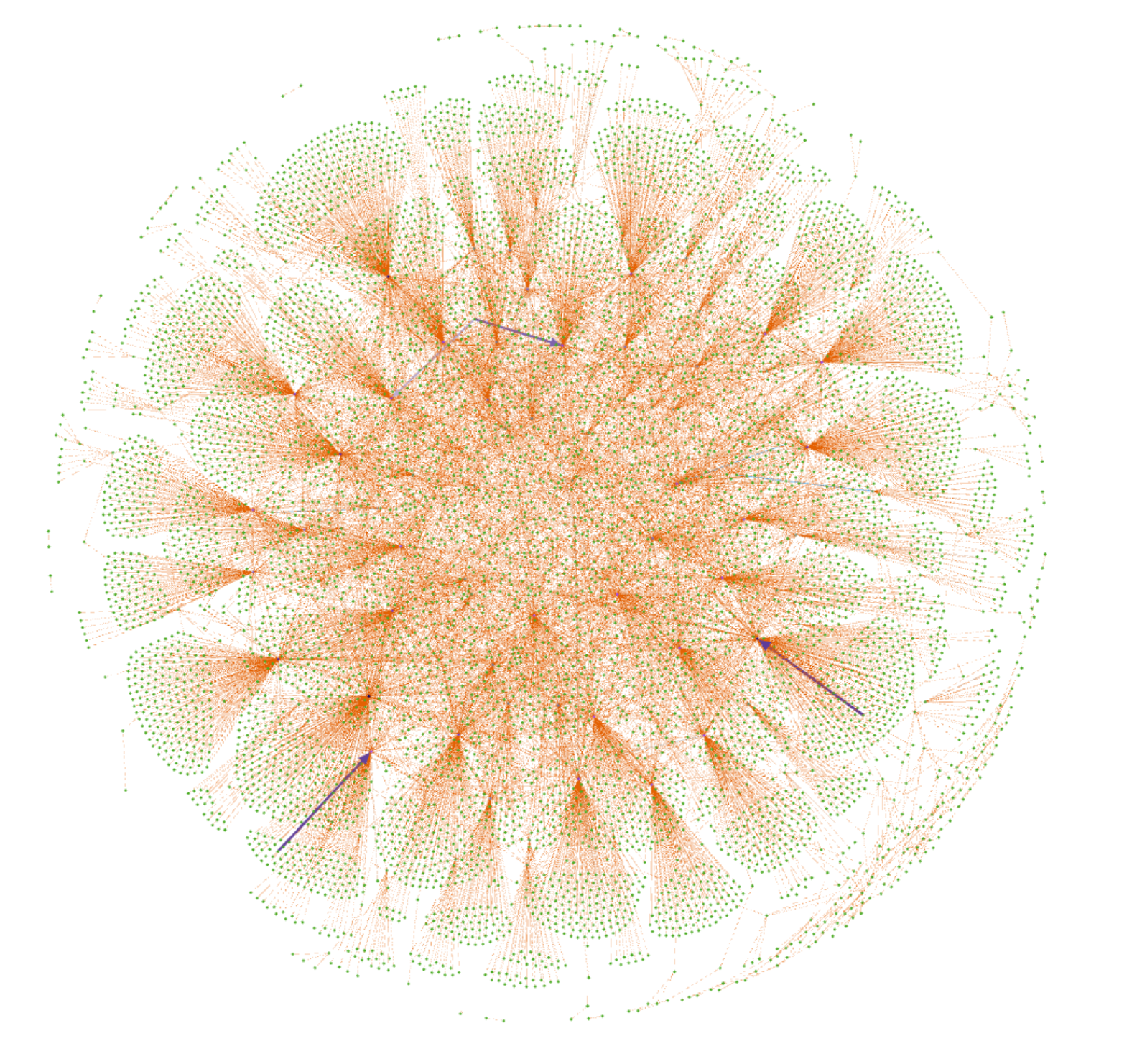}
	}
	\caption{The visualization of AVG.}
	\label{fig:AVGVisualization}
	\end{figure*}

	For AVG, there are 28,769 nodes and 439,154 edges, and the total weight value of all edges in the graph is 1,519,464, which indicates that some voters often vote for the same producers in different actions. One possible explanation for this phenomenon is that some candidates are very trustworthy and have their faithful supporters.
	\begin{figure}[htbp]
	\centering
	\subfigure[Degree distribution]{%
		\includegraphics[scale=0.28]{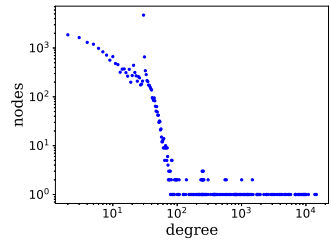}
	}
	
	\subfigure[Indegree distribution]{%
		\includegraphics[scale=0.28]{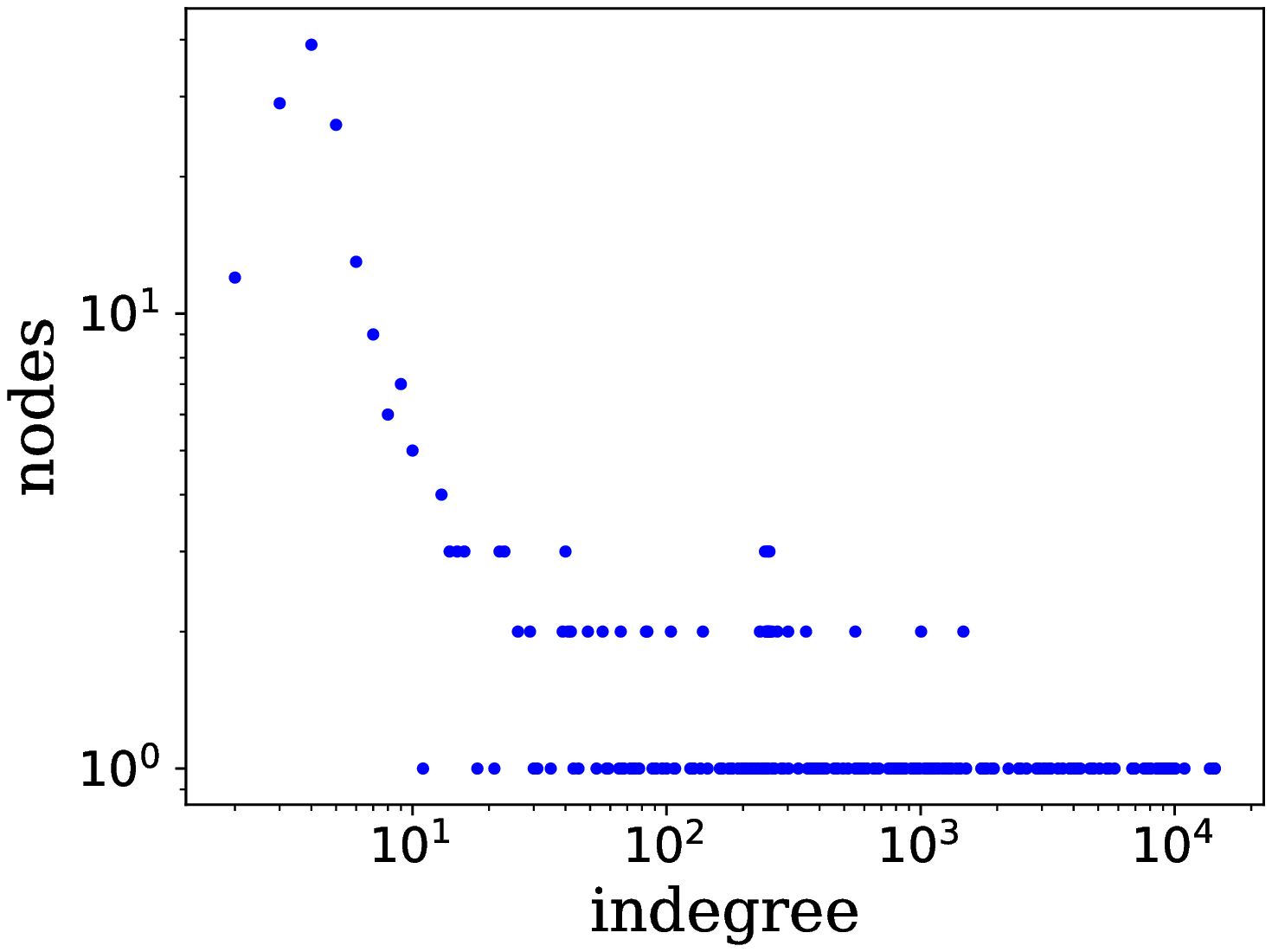}
	}
	\subfigure[Outdegree distribution]{%
		\includegraphics[scale=0.28]{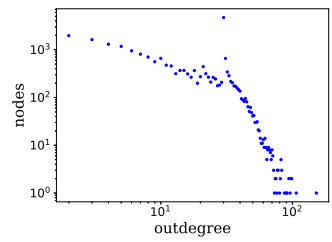}
	}
	\centering
	\caption{ Degree/Indegree/Outdegree distributions of AVG}
	\label{fig:AVGdegree}
	\end{figure}
	
	Fig. \ref{fig:AVGVisualization} displays the visualization result of AVG, which contains 10,000 edges selected from AVG. The thickness of each edge is proportional to its weight value. We observe that several edges are obviously thicker than others, which account for the continued supports to some candidates. There exist some hub nodes in the graph, representing some influential candidates. Besides, many edges are randomly distributed in AVG, which reflects that in the early EOSIO, some voters participate in voting for testing or experiencing.
	
	
	As shown in Fig. \ref{fig:AVGdegree}, the degree distribution, indegree distribution and outdegree distribution of AVG can not strictly satisfy the power law distribution. We can observe that candidates with a large number of supporters occupy a small proportion of all candidates. And the voting times for most voters are few, which may be related to the rule that voters should mortgage a part of EOS tokens when voting.
	
	
	
    The results of some graph metrics for AVG are shown in Table \ref{table:Metrics of Graphs}. As we can see, the clustering coefficient is 0.066, namely there are few triangles in AVG. If a voter votes for two candidates, these two candidates will barely vote for each other owing to their competitive relationship. The assortativity coefficient is negative, which implies that large-degree nodes tend to vote for or be voted by small-degree nodes. The number of nodes in the largest SCC is 18, indicating that there exist some few voting gangs in which the members vote for each other. More than 99.98\% nodes in AVG participate in the largest WCC, which means that accounts are almost fully connected in AVG, and this phenomenon is similar in some other complex networks \cite{Cha2010} \cite{leskovec2007graph}. The largest diameter of WCC is 6, meaning that the distance between two nodes in AVG is small. Besides, none of the candidates vote for themselves, which is reflected by the smallest diameter.
	
	
	
	
	\subsection{Money Transfer}
	
	We construct a money transfer graph (MTG) to investigate the transfer relationship between accounts as follow:
	
%
%
	\begin{myDef}[MTG] A money transfer graph (MTG) is a directed weighted graph $G=(V,E,W)$ with a mapping function ${\varphi}: E\rightarrow W$ that maps a weight from the edge attribute set $W$ for each edge, where $V$ denotes the set of nodes which represent accounts participating in transferring EOS, $E$ denotes the set of edges, in which each edge $(v_i,v_j, w)$, $v_i,v_j\in V, w\in W$ represents that $v_i$ totally transfers $w$ EOS to $v_j$.
	\end{myDef}

	\begin{figure*}[htbp]
		\centerline{
			\includegraphics[scale=0.15, trim= 0 -100 0 0]{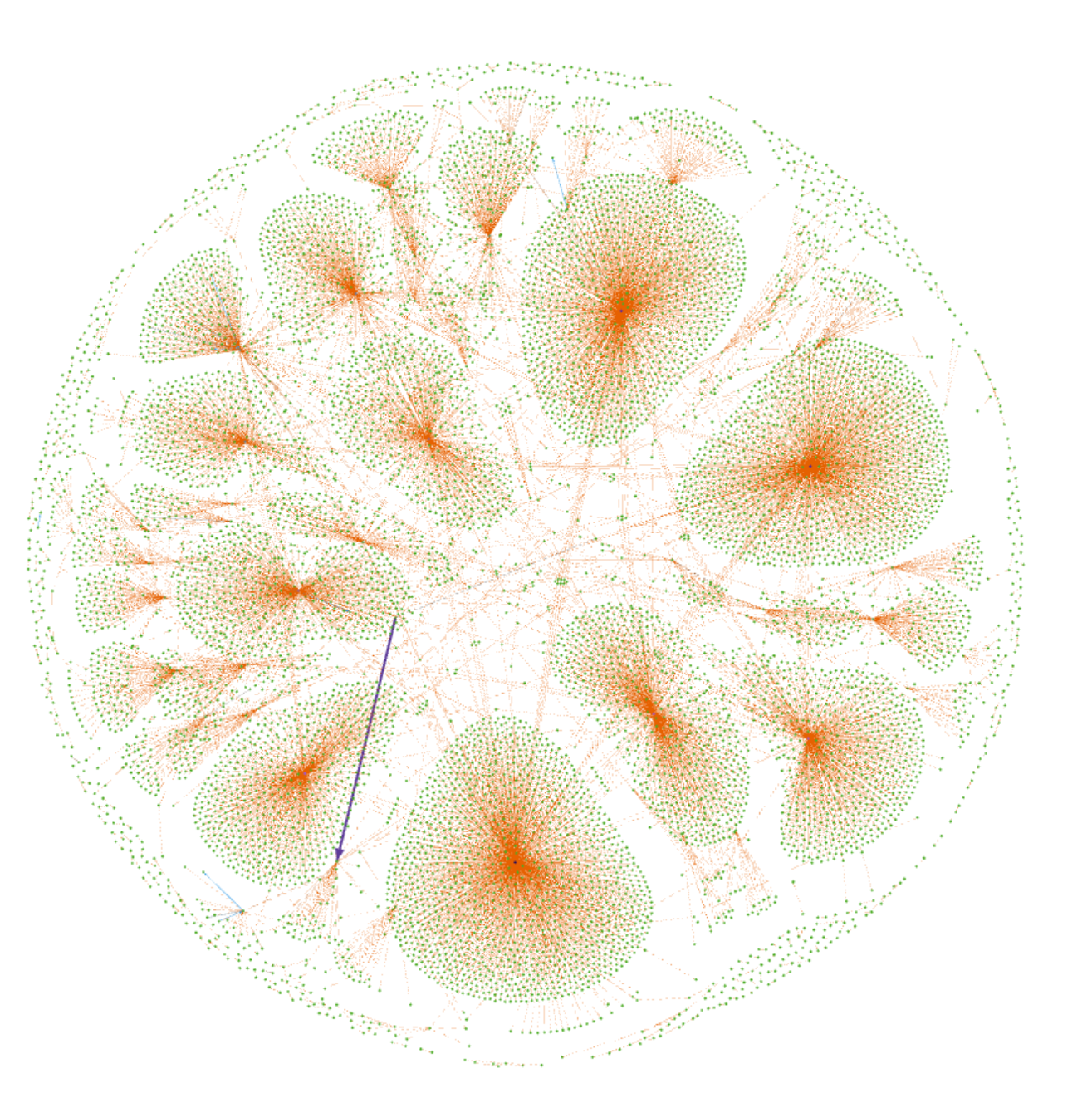}
		}
		\caption{The visualization of MTG.}
		\label{fig:MTGVisualization}
	\end{figure*}

	MTG contains 204,841 nodes and 1,370,813 edges in total. According to Table \ref{Table:Proportion}, there are 8,622,288 actions related to money transfer, which is about 6 times the edge number. That is, there exist repeated money transfer actions between two accounts.
	
	We visualize MTG by sampling 10,000 edges from MTG, and make the thickness of each edge be proportional to its weight value. As shown in Fig. \ref{fig:MTGVisualization}, the sampling subgraph of MTG exist significant community structures. It contains a few large-degree nodes, which are community centers and interact frequently with surrounding nodes, these nodes may be exchanges or accounts of some DApps in charge of their ledger. There are also a large number of small-degree nodes in MTG, the free for usage feature offers a low-barrier entry point for individual users.
	
	\begin{figure}[htbp]
		\centering
		\subfigure[Degree distribution]{%
			\includegraphics[scale=0.28]{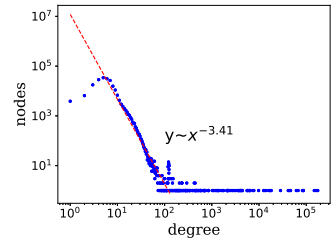}
		}
	
		\subfigure[Indegree distribution]{%
			\includegraphics[scale=0.28]{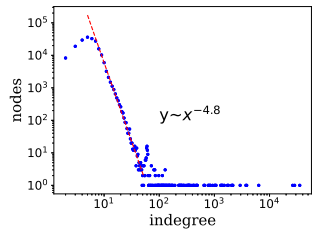}
		}
		\subfigure[Outdegree distribution]{%
			\includegraphics[scale=0.28]{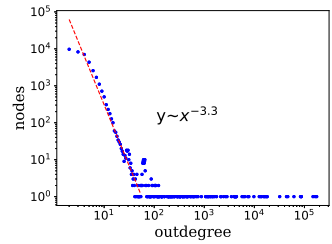}
		}
		\centering
		\caption{ Degree/Indegree/Outdegree distributions of MTG}
		\label{fig:MTGdegree}
	\end{figure}
	
	The degree distribution, indegree distribution and outdegree distribution are shown in Fig. \ref{fig:MTGdegree}. We can observe that these distributions follow the power law, meaning that there are a few accounts take part into money transfer activities for many times, most of the accounts are small-degree nodes. By investigating the fitting line $y\sim x^\alpha$ we plotting, we can draw a conclusion that the distribution of outdegree is more variable than the distribution of degree and indegree, since the larger the $\alpha$, the more variable of the degree.
	
	As shown in Table \ref{table:Metrics of Graphs}, the clustering coefficient of MTG is 0.259, which is a relatively large value, indicating that if an account has transactions with other two accounts respectively, these two accounts tend to have money transferring relationships.
	The assortativity coefficient of MTG is negative, revealing that large-degree nodes tend to connect to small-degree nodes in MTG. The number of nodes in the largest SCC is 55,238, which accounts for about 26.97\% of all the nodes in MTG. It indicates that there exist some hub nodes (e.g., system accounts, exchanges) or sham transactions that could cause large SCC in MTG. The largest WCC contains all nodes in MTG, since the EOS tokens are initially managed by the EOSIO's system accounts, and then they are distributed to the wallet of individual users by a series of processes. The diameter of MTG is small and the clustering coefficient is large, indicating a small world phenomenon in MTG.
	
	\subsection{Contract Authorization}
	According to the design of EOSIO, it is hard to know which account invokes the contract for a contract invocation. However, the information that which account delegate its permission for the execution of a contract can be extracted from the action trace data. We construct a contract authorization graph (CAG) to describe relationships in contract authorization activity. The definition is shown as follow:
%
	\begin{myDef}[CAG] A contract authorization graph (CAG) is a directed weighted graph $G=(V,E,W)$ with a mapping function ${\varphi}: E\rightarrow W$ that maps a weight from the edge attribute set $W$ for each edge, where $V$ is the set of nodes representing accounts taking part in the contract authorization activity, $E$ is the set of edges, in which each edge $(v_i,v_j, w)$, $v_i,v_j\in V, w\in W$ represents that $v_i$ delegates its permissions to execute the contract of $v_j$ for $w$ times.
	\end{myDef}

\begin{figure*}[htbp]
	\centerline{
		\includegraphics[scale=0.12,trim=20 20 20 10]{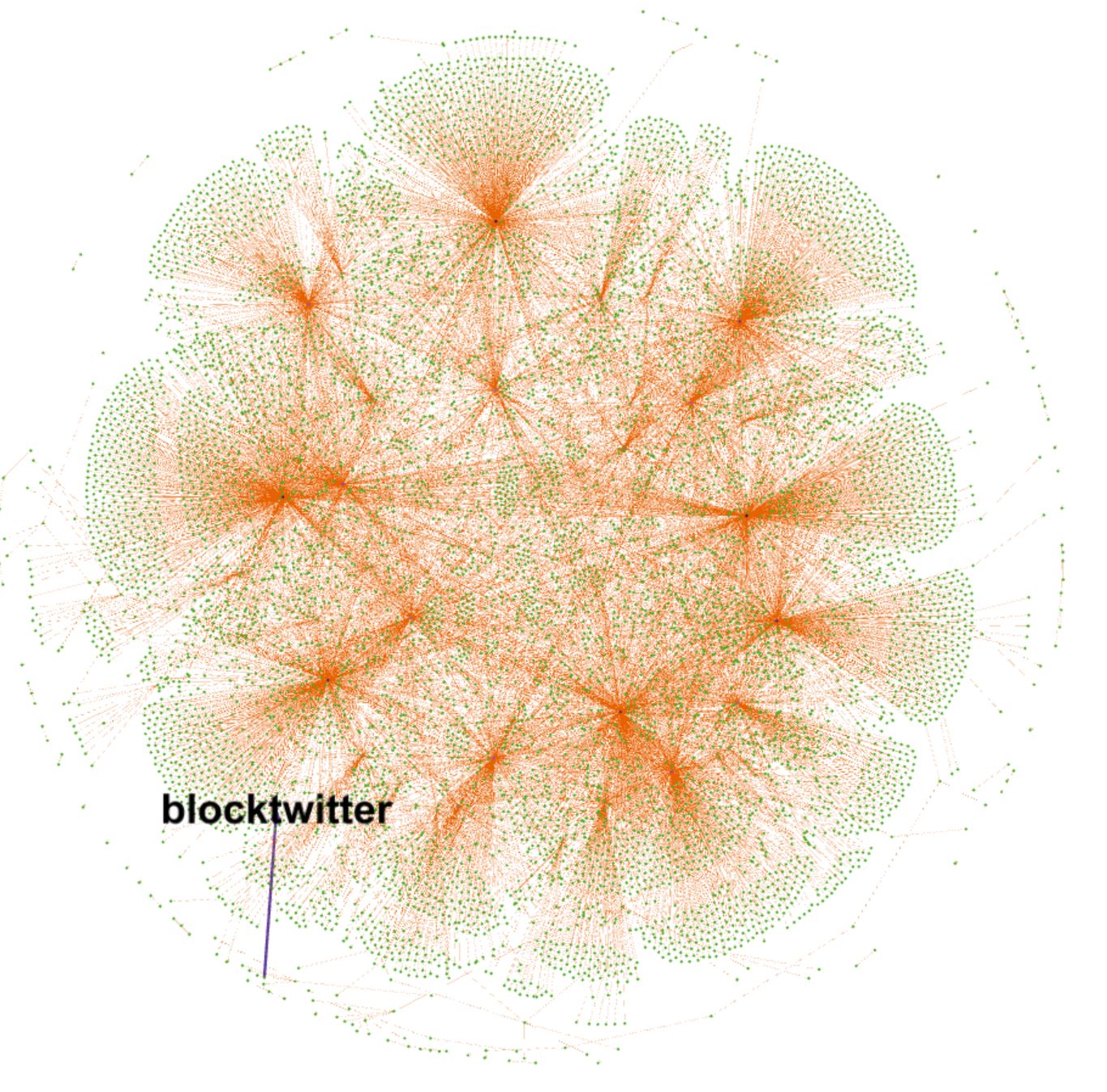}
	}
	\caption{The visualization of CAG.}
	\label{fig:CAGVisualization}
\end{figure*}	

	There are 35,479 nodes and 126,918 edges in CAG. The total weight value of all edges in the graph is 331,530,705. However, we observe that the total weight value of edges belonging to an account ``blocktwitter'' is 316,579,248, which occupies a major part of the total weight in the graph. The contract of ``blocktwitter'' is reported to be an abnormal account that would periodically launch a great many of actions named tweets for pressure testing \cite{zheng2020xblock}, and it behaves like a contract for denial of service attack.
	
	We visualize CAG by sampling 10,000 edges from CAG. The visualization result is shown in Fig. \ref{fig:CAGVisualization}, where the thickness of each edge is proportional to its weight value. As we can see in the figure, there exist some nodes with large degree in CAG, it may because some influential contracts (e.g. contracts of some popular DApps) are invoked by many accounts, or some active accounts invoke a large number of different contracts. Besides, an edge of ``blocktwitter'' is conspicuous in the graph with large thickness, which is consistent with our analysis.
	
	\begin{figure}[htbp]
		\centering
		\subfigure[Degree distribution]{%
			\includegraphics[scale=0.28]{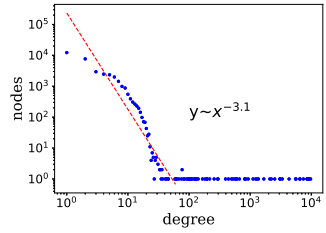}
		}
	
		\subfigure[Indegree distribution]{%
			\includegraphics[scale=0.28]{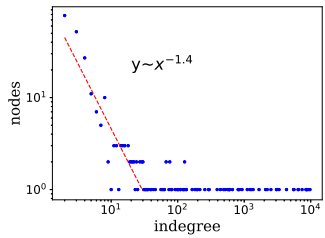}
		}
	    \subfigure[Outdegree distribution]{%
			\includegraphics[scale=0.28]{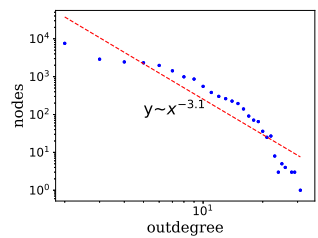}
		}
		\centering
		\caption{Degree/Indegree/Outdegree distributions of CAG}
		\label{fig:CAGdegree}
	\end{figure}

	Fig. \ref{fig:CAGdegree} shows the degree distribution, indegree distribution and outdegree distribution of CAG. All these distributions approximately follow the power law. From the indegree distribution, we can know that few contracts have been invoked for many times. That is, not all contracts are widely known and invoked by users. For the outdegree distribution, few accounts delegate their permissions to others for many times. Besides, there exists no long tail in the outdegree distribution, and the outdegree gap is small, which means that many users in the early EOSIO only interact with a limited number of contracts. Of course, this phenomenon may also be related to the small number of smart contracts in the initial phase of EOSIO.
	
	
	Results in terms of several graph metrics for CAG are shown in Table \ref{table:Metrics of Graphs}. As we can see, the clustering coefficient is 0.086, which means if an account invokes two contracts respectively, these two contracts will barely invoke each other.
	The assortativity coefficient is negative, illustrating that large-degree nodes tend to interact with small-degree nodes in contract authorization activity. The number of nodes in the largest SCC is 3, which implies that there may exist some accounts belonging to the same owner (e.g., a DApp), and their contracts collaborate closely to complete some specific functions. More than 98.89\% nodes are contained in the largest WCC. The smallest diameter of WCC is 0 since some accounts invoke their own contract.
	
	\section{Conclusion} \label{Conclusion}
	In this paper, we conducted a comprehensive graph analysis on EOSIO by considering four main activities in EOSIO, including account creation, account vote, money transfer and contract authorization. Via utilizing both the on-chain data and off-chain data, we constructed four graphs for these activities. We then characterized these graphs using several complex network metrics, and obtained some interesting insights and observations such as there exist some few gangs in which the members vote for each other, which helps people have a deep understanding on EOSIO. For future work, we will focus on some abnormal behaviors revealed in our study, and provide constructive suggestions for EOSIO supervision.
	%
	%
	%
	
	\bibliographystyle{splncs04}
	\bibliography{mybib}
	
\end{document}